\documentclass[12pt]{iopart}

\usepackage{graphicx}
\usepackage{graphics}
\usepackage{dcolumn}
\usepackage{bm}
\usepackage{color}
\usepackage{soul}
\usepackage{textcomp}
 \usepackage{multirow}
\usepackage{soul}
\usepackage{rotating}
\usepackage{setspace}
\usepackage[mathlines]{lineno}
\usepackage{enumerate}
\usepackage{float}
\usepackage{blindtext}
\usepackage{subfigure}
\usepackage{microtype}
\usepackage{amssymb}

\usepackage{soul}
\usepackage[breaklinks=true,hyperindex=true,
pdftitle={Stealth Chaos due to Frame-Dragging},
colorlinks=true,pagebackref=false,citecolor=blue,plainpages=false,
pdfpagelabels,
linkcolor=blue,urlcolor=blue]{hyperref}

\def\XXint#1#2#3{{\setbox0=\hbox{$#1{#2#3}{\int}$}
\vcenter{\hbox{$#2#3$}}\kern-.5\wd0}}

\newcommand{\GFTyMA}{
Grupo de F{\'i}sica Te{\'o}rica y Matem{\'a}tica Aplicada, 
Instituto de F{\'i}sica, 
Facultad de Ciencias Exactas y Naturales, 
Universidad de Antioquia, Medell\'in, Colombia.}
\newcommand{\Konrad}{
Programa de Matem\'atica, 
Fundaci\'on Universitaria Konrad Lorenz, 110231 Bogot\'a, Colombia.}
\newcommand{\UIUC}{
Illinois Center for Advanced Studies of the Universe, 
Department of Physics, University of Illinois at Urbana-Champaign, 
Urbana, IL 61801, USA.}

\begin{document}
\title{Stealth Chaos due to Frame-Dragging}

\author{%
Andr\'es F. Gutierrez$^{1}$
Alejandro~C\'ardenas-Avenda\~no$^{2}$ $^{3}$, 
Nicol\'as~Yunes$^{3}$ and
Leonardo A. Pach\'on$^{1}$
}

\address{$^{1}$~\GFTyMA}
\address{$^{2}$~\Konrad}
\address{$^{3}$~\UIUC}

\begin{indented}
\item[]\date{\today}
\end{indented}

\begin{abstract}

While the geodesic motion around a Kerr black hole is fully
integrable and therefore regular, numerical explorations of the phase-space
around other spacetime configurations, with a different multipolar
structure, have displayed chaotic features. 
In most of these solutions, the role of the purely general relativistic
effect of frame-dragging in chaos has eluded a definite answer. 
In this work, we show that, when considering neutral test particles around 
a family of stationary axially-symmetric analytical exact solution to the Einstein-Maxwell field equations, frame-dragging (as captured through spacetime vorticity)is capable of reconstructing KAM-tori from initially highly chaotic configurations.
We study this reconstruction by isolating the contribution of
the spacetime vorticity scalar to the dynamics, and exemplify our findings by computing
rotation curves and the dimensionality of phase-space. 
Since signatures of chaotic dynamics in gravitational waves have
been suggested as a way to test general relativity in the strong-field
regime, we have computed gravitational waveforms using the semi-relativistic
approximation and studied the frequency content of the gravitational wave
spectrum.
If this mechanism is generic, chaos suppression by frame-dragging may undermine present proposals to
verify the hypothesis of the no-hair theorems and the validity of
general relativity.
\end{abstract}

\maketitle

\section{Introduction}
One of the most studied sources of gravitational waves are 
extreme-mass-ratio inspirals (EMRIs), which consist of a low mass compact
object, like a stellar black hole or a neutron star that inspirals into a
supermassive black 
hole~\cite{Babak:2017tow,Gair:2017ynp,Barausse:2020rsu}.
The gravitational waves emitted by these types of systems constitute a way
to map the multipole structure of the spacetime, i.e., if the spacetime is
stationary and axially-symmetric, the gravitational waves generated by a
particle in a near-circular orbit, carry the multipole moment information
of the central object~\cite{Ryan:1995wh,Ryan:1995xi}. 
If the central object is a Kerr black hole, the no-hair
theorems~\cite{Israel:1966rt,Carter:1971zc,Hawking:1971vc} for
electrically-neutral, isolated black holes imply that all the multipole
moments of the source are uniquely and fully characterized by the mass and
the spin angular momentum of the source.  
Due to the particular features of the Kerr solution, evidence of black
holes outside general relativity, containing higher-order independent
multipole moments, would lead to a change of paradigm in gravitational
physics~\cite{Barack:2006pq,Barausse:2006vt,Gair:2007kr,Barausse:2008xv,
Apostolatos:2009vu,LETELIER20113655,Brink:2013nna,
Cardenas-Avendano:2018ocb,Datta:2019euh}.  

In terms of dynamical systems, the geodesic motion around a Kerr black
hole is fully integrable and can be solved by the specification of four
constants of motion, namely the rest mass of the orbiting body, the orbital
energy, the orbital angular momentum and the Carter constant, $M$, $E$, $L$
and $Q$, respectively~\cite{Carter:1968ks,Schmidt:2002qk,Teukolsky:2014vca}. 
Under a Hamiltonian formulation of the geodesic motion,
chaos refers to the non-integrability of the equations of motion, i.e.,
the non-existence of a smooth analytic function that interpolates between
orbits~\cite{Levin:2006zv}. 
This notion of integrability of Hamiltonian systems is known 
as Liouville integrability: if $n$ linearly independent integrals of motion
exist in a system of $n$ degrees of freedom, then there exists a coordinate
transformation to angle-action variables such that the equations of motion
can be put in quadrature form~\cite{Contopoulos_2002}.

According to the Kolmogorov–Arnold–Moser (KAM) theorem~\cite{tabor1989chaos},
small perturbations of an integrable system imply that some of the invariant
tori are deformed and survive, while others are destroyed. 
These perturbations can arise from external sources (e.g., discs
or rings~\cite{Semerak:2012dw,Polcar:2019wfi} surrounding the source) or due
internal changes in the multipole structure of the source; the latter are the
ones we explore in this work.

When the perturbation destroys the invariant tori, the resonant
orbits in phase space disintegrate to form Birkhoff chains of
islands~\cite{rasband2015chaotic}, where the fundamental orbital frequencies
are different for each member of the chain, although the ratio of these
frequencies is the same for each curve~\cite{Apostolatos:2009vu}. 
In contrast, the variation of the frequency between consecutive orbits,
which do not belong to a Birkhoff island or a chaotic
layer~\cite{Contopoulos_2002,Apostolatos:2009vu}, is a monotonic function
of the radial coordinate. 
This unique characteristic can be used as an indicator of extra multipole
moments of the central source and the observation of a gravitational wave
with a constant frequency ratio could imply the presence of a non-Kerr
background~\cite{Apostolatos:2009vu}. 

Over the past years, several numerical explorations of the phase-space
around astrophysical sources that have an independent quadrupole moment
have shown the appearance of chaos in exact solutions to the Einstein
field equations that have arbitrary multipole moments. 
For example, the Manko-Novikov~\cite{1992CQGra...9.2477M} and the
Pachon-Rueda-Sanabria-Gomez (PRS)~\cite{Pachon:2006an,Ruiz:2015qla}
solutions are axially symmetric configurations that guarantee that the
mass $M$, $E$ and $L$ are constants of the motion but they do not possess
an analogous Carter constant~\cite{Brink:2008xy, Han:2008zzd,
Brink:2009mq, Contopoulos:2011dz}, and therefore, geodesic chaos is
expected. 

With the future space-based detector Laser Interferometer Space
Antenna (LISA) it may be possible to measure fractional deviations from the
quadrupole moment of the Kerr solution at a sub-percent level of
accuracy~\cite{Barack:2006pq,Barausse:2020rsu}. 
Thus, depending on the mass ratio, there could be EMRIs that will slowly
cross a resonance and, if the spacetime is not regular, leave a detectable
signature of chaos~\cite{Apostolatos:2009vu}.

In this paper, we show that, for neutral particles, the introduction of the
purely general relativistic effect of frame-dragging, as parameterized through a vorticity scalar computed from the vorticity tensor of a congruence  world-lines describing static observers~\cite{Misner:1974qy}, is capable of reconstructing the KAM tori from an initial highly chaotic configuration
around a stationary gravitational source  described by the PRS metric~\cite{Pachon:2006an}. Given that the vorticity scalar encodes the precession frequency of a gyroscope attached to the world-lines of a static observer, this quantity is a good proxy for frame-dragging effects.
If this effect is generic, frame-dragging may suppress chaotic signatures e.g.,
Birkhoff chains in EMRIs around non-Kerr objects, which would prevent probes
of the assumptions of the no-hair theorems with gravitational waves. 

The layout of this paper is as follows. 
In Sec.~\ref{metricanddefs} we introduce the PRS metric and the figures of
merit we use to display the presence of chaos. 
To exemplify the effect of rotationally induced multipole
moments on the EMRI dynamics around a central compact object, in
Secs.~\ref{KAMI} and~\ref{KAMII}, respectively, we study the following two
relevant scenarios: in \emph{KAM Tori Reconstruction I}, we consider a
stationary central object without an electromagnetic field; and in \emph{KAM Tori Reconstruction II}, we consider a stationary central object with an electromagnetic field. 
In both of these examples, the orbiting neutral particle develops a mixture of regular and
chaotic dynamics, but as the vorticiy of the spacetime is increased, the
chaotic features are diminished. 
In Sec.~\ref{structure} we provide a theoretical explanation of why chaos is being suppressed in this context. 
Finally, in Sec.~\ref{Future}, we present our discussion, ideas for future
work and conclude.
Throughout this paper we use geometric units with $c=1=G$.

\section{Theoretical Framework}
\label{metricanddefs}
The emergence of chaos around astrophysical objects that deviate from the
Kerr black hole is driven by the intricate interplay between strong field
corrections from different multipole orders. 
Let us then consider the PRS metric, designed to describe the exterior
gravitational field of a reflection-symmetric source, with generic mass monopole $M$, spin angular momentum per unit mass $a=J / M$, gravitational
quadrupole moment $\mathcal{Q}^{\rm{G}}$, current octupole moment
$\mathcal{S}^{\rm{G}}$ and mass hexadecapole moment
$\mathcal{H}^{\rm{G}}$, along the symmetry axis. 
This solution is given in terms of the quasi-cylindrical
Weyl-Lewis-Papapetrou coordinates $x^{\mu} = (t,\rho,z,\phi)$ as
\begin{equation}
\label{Papapetrou} 
\mathrm{d} s^2=-f(\mathrm{d}t-\omega \mathrm{d}\phi)^2 +
f^{-1}
\left[e^{2\gamma} (\mathrm{d} \rho^2 
+ \mathrm{d} z^2)+\rho^2 \mathrm{d} \phi^2\right],
\end{equation}
where the metric functions $f(\rho,z)$, $\omega(\rho,z)$ and $\gamma(\rho,z)$
are obtained from the Ernst complex potentials ${\cal E}(\rho,z)$ and
$\Phi(\rho,z)$, which obey the following relations~\cite{Ernst:1967by}
\begin{eqnarray}
(\Re \left[{\cal E}\right]+|\Phi|^2)\nabla^2{\cal E}
&=& (\nabla {\cal E} + 2\Phi^*\nabla\Phi)\cdot\nabla{\cal E}, 
\nonumber \\ 
(\Re\left[{\cal E}\right]+|\Phi|^2)\nabla^2\Phi&=& (\nabla {\cal E} +
2\Phi^*\nabla\Phi)\cdot\nabla\Phi\, . 
\label{Ernst}
\end{eqnarray}

The Ernst equations (\ref{Ernst}) can be solved by means of the
Sibgatullin's integral method~\cite{Manko_1993}, according to which 
the complex potentials $\cal E$ and $\Phi$ can be calculated from
specified axis data ${\cal E}(z,\rho=0)$ and $\Phi(z,\rho=0)$. 
The details and the explicit solution can be found in
Refs.~\cite{Pachon:2006an,Ruiz:2015qla}. 
Here we will only show the multipole moments, which are~\cite{Pachon:2006an}
\begin{eqnarray}
\label{multipolosP3}
\mathcal{Q}^{\rm{G}} &=&   M(k-a^2),  
\qquad 
\mathcal{S}^{\rm{G}} = - Ma(a^2 - 2 k ),
\nonumber \\
\mathcal{H}^{\rm{G}} &=&
\frac{1}{70} M \left[70 a^4-210 a^2 k + 13 a q \mu + 
10 k (7 k-M^2+q^2)+3\mu^2\right]. 
\end{eqnarray}
The parameter $k$ allows the source to have an arbitrary mass quadrupole
moment, without changing its mass or spin angular momentum, while the
magnetic dipole parameter $\mu$ changes the mass hexadecapole moment,
without changing any other lower multipole. 
In the electromagnetic sector, the solution has a magnetic dipole moment
$\mathcal{B}^{\rm{E\&M}} =  \mu + aq$ and an electric quadruple moment
$\mathcal{Q}^{\rm{E\&M}} = - a^2 q -a \mu +kq$, where $q$ is the electric
charge.
The electromagnetic field is then fully determined by the charge $q$, the
parameter $\mu$, the dimensional spin angular momentum $a$ and the
gravitational mass quadrupole parameter $k$. 
Observe that one can have an electric field even when $q=0$ that is
induced by rotation, i.e., provided $\mu$ and $a$ are 
non-vanishing~\cite{Herrera:2006cw}. 
The PRS metric reduces to the Kerr-Newman solution when $k=0$ and 
$\mu = 0$. 

In the PRS solution, the vorticity scalar of spacetime $\omega_{\mathrm{v}}$ originates
from a combination of the mass currents and the non-vanishing Poynting vector
of the electromagnetic field, which can be written as~\cite{Ruiz:2015qla}
\begin{equation}
 \omega_{\mathrm{v}}=\sqrt{\omega^\alpha{}_\beta \omega_\alpha{}^\beta}=\frac{e^{-\gamma}}{\sqrt{2f}}\sqrt{
 \Im[\mathcal{E}_{,z} + 2\Phi^{*}\Phi_{,z}]^{2} +
 \Im[\mathcal{E}_{,\rho}+2\Phi^{*}\Phi_{,\rho}]^{2} },
  \label{equ:vorticity}
\end{equation}
where the vorticity tensor is defined as $\omega_{\alpha\beta}=u_{\left[\alpha;\beta\right]}+\dot{u}_{\left[\alpha\right.}u_{\left.\beta\right]}$ for a congruence of static observers with tangent vector $u^{\alpha}$~\cite{Herrera:2000uh}. The vorticity scalar is a function of global quantities of the spacetime, such us the spin angular momentum, deformation parameters, or the charge for the electromagnetic case. One can show, as done by Rindler and others in the 1990s~\cite{rindler}, that the vorticity scalar is directly proportional to the rate of precession of gyroscopes outside of a massive source. It is in this sense that the vorticity scalar encodes frame dragging effects.

\subsection{Figures of merit: the rotation number and the correlation
dimension}
We quantitatively study the possible reconstruction of KAM tori due to
frame-dragging through rotation curves $\nu_{\theta}$~\cite{Contopoulos_2002}
and the dimensionality $D$ of phase space through the correlation dimension
$C(\varrho)$~\cite{Grassberger:1983zz}. 
The rotation curve $\nu_{\theta}$ provides information on the localization
and emergence of Birkhoff islands, while the correlation dimension encodes
the dimensionality of the manifold in which the initial condition is
allowed to evolve by considering correlations between points of a
long-time series on the manifold~\cite{Grassberger:1983zz}.

The rotation number is defined as~\cite{Contopoulos_2002} 
\begin{equation}
\nu(\rho) = 
\lim_{N \rightarrow \infty} \frac{1}{2\pi N}\sum_{i=1}^{N}\theta_i,
\label{Eq:tornum}
\end{equation}
where $\theta_i$ is the clockwise angle subtended by two vectors, defined
from the invariant point, $\left(\rho, \dot{\rho} \right)$, to two
consecutive successive piercings of the Poincare's surface of section,
i.e., $\theta_i = \measuredangle 
(\vec{v}_{i+1},\vec{v}_{i})$~\cite{Contopoulos_2002}.
The rotation curve of the system is obtained by evaluating the rotation
number as a function of the location of the Poincar\'e surface of section
in phase space, which for this work we define as the location of the
surface by the minimum value of the radial coordinate sampled by that 
surface. 
Abrupt changes in the rotation curve signal the presence of chaotic
orbits~\cite{Contopoulos_2002,Apostolatos:2009vu}.

The chaotic features displayed in the rotation curve can sometimes be hard to
discern by naked eye, due to subtle changes in the monotonicity of the curve.
As an alternative, it is possible to focus on the dimension of
the invariant manifold in which the motion takes place, as conservative chaotic systems have an integer dimension of the invariant manifolds, whereas the dissipative have a non-integer dimension~\cite{LR03}.

Among different proposals for measuring the invariant-manifold dimension
\cite{LR03}, we focused on the correlation dimension introduced by
Grassberger \& Procaccia in Ref.~\cite{Grassberger:1983zz}. In this approach, the phase-space distance $\varrho\ll1$ and the correlation dimension
$C\left(\varrho \right)$ is proportional to $\varrho^{D}$, where $D$ is the
dimensionality of of the phase-space orbit.
In operational terms, $C\left(\varrho \right)$ is calculated
as follows.
Given $N$ sampled points on the orbit, the correlation dimension is defined
as~\cite{Barnes_2001}
\begin{equation}
C\left(\varrho \right)=\lim_{N\rightarrow\infty}\frac{1}{N^{2}}
\sum_{i=1}^{N}\sum_{j=1,j \neq i}^{N} 
\Theta\left(\varrho-\left|\xi_{i}-\xi_{j}\right|\right),
\label{Eq:corr_int}
\end{equation}
where $\xi_{i}=\left( \rho, \dot{\rho}, z, \dot{z} \right)$ represents a
phase-space vector and $\Theta$ is the Heaviside step function. 
Orbits with $D = 1$ are close to the invariant point or to small
Bifkhoff islands that surround periodic orbits, while orbits with 
$D = 2$ or $3$ are integrable \cite{LR03}.

With the background defined and these two tools to study the
dynamics, we now follow the procedure described in
Ref.~\cite{Cardenas-Avendano:2018ocb} to obtain the geodesic motion in the
PRS metric. 
We numerically integrate the Hamiltonian equations of motion for the
variables $\dot{\rho},\dot{z},\dot{P_{\rho}},\dot{P_{z}}$, using an explicit
Runge$-$Kutta method due to Dormand and Prince~\cite{DORMAND198019}. 
We have checked that the conserved quantities remain constant, e.g.,
errors did not exceed the order of $\Delta E / E = 10^{-12}$,  under the
numerical evolution, ensuring that the particle does not wander in phase
space due to numerical error.

\section{KAM Tori Reconstruction I}
\label{KAMI}
Consider, as an example, a stationary compact object described by the PRS metric~\cite{Pachon:2006an,Ruiz:2015qla} with mass $M$, mass
quadrupole moment  $\mathcal{Q}^{\rm{G}}=1.6 \, M^3$, and $j=0=\mu$,
implying $k \neq 0$. 
Here we have introduced the dimensionless spin parameter $j \equiv a/M$.
Since $\mathcal{Q}^{\rm{G}}>0$ when $j = 0$, the source is prolate with a
\emph{significant} deviation from the Kerr intrinsic oblate deformation,
and therefore, the corresponding phase space will suffer a substantial
deviation from integrability, and one expects chaotic
dynamics~\cite{LETELIER20113655}.

\begin{figure*}[htb]
\includegraphics[width=\columnwidth,clip=true]{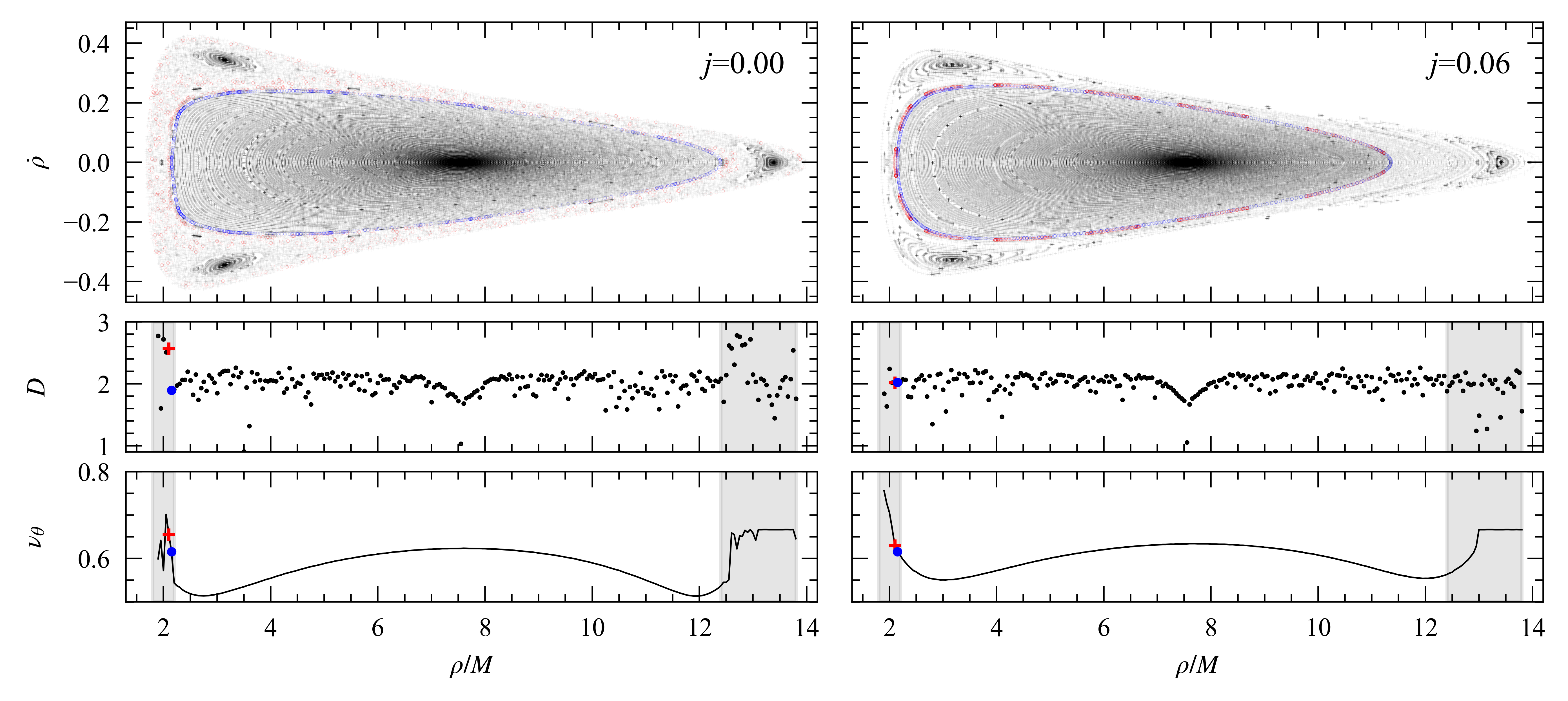}
\vspace{-0.3cm}
\caption{Poincar\'e surfaces of section (top), the dimensionality of the
phase-space orbits (middle), and the rotation curves (bottom), for orbits
with $E=0.95$,  $L=3\,M$ and $\mathcal{Q}^{\rm{G}}=1.6\,M^3$. 
The left panels correspond to a background with zero dimensionless spin
parameter $j=0.0$, while the right panels have $j=0.06$,  while
maintaining $\mathcal{Q}^{\rm{G}}$ constant. 
The increase in the dimensionless spin parameter decreases the chaotic
structure observed in the rotation curves and in the dimensionality,
especially in the shaded regions. 
The two middle and lower panels also highlight the two particular initial
conditions (one at $\rho=2.10\, M$ (red) and one at $\rho=2.15\, M$
(blue)) for which we will calculate gravitational waves in the
semi-relativistic approximation. 
}
\label{fig:quadrupoleIC}
\end{figure*}

The top-left panel in Fig.~\ref{fig:quadrupoleIC} shows the Poincaré
surfaces of section for a test particle moving on this background with
$E=0.95$ and $L=3\,M$. 
The main islands in the center of the figure are surrounded by a chaotic
sea of layers with many high-multiplicity islands of stability, caused by
geodesics that visit the same equilibrium point multiple times. 
This strong chaos is also manifest in the associated dimensionality of
phase space and the rotation curves, shown in the middle and bottom panels
of the figure, respectively. 
Observe how in the shaded regions, $\nu_{\theta}$ shows non-monotonic
variations and a large plateau, all of which are clear signatures of a
non-integrable system. 

Let us now endow the spacetime with vorticity, and therefore
introduce frame-dragging effects, by increasing the spin angular momentum
of the source to a fixed number, while keeping $\mu=0=q$, and
varying $k$ in Eq.~(\ref{multipolosP3}) in such a way that
$\mathcal{Q}^{\rm{G}}$ remains constant. 
We find that when doing so, the integrability of the orbits increases, in
the sense that the rotation number becomes more regular and the
dimensionality of the phase-space manifold goes down to two. 
The right panels of Fig.~\ref{fig:quadrupoleIC} show this result for
orbits with the same initial conditions as the left panel, 
but with $j=0.06$ (equivalent to an angular rotation frequency of the
horizon of a few times $10^{-3}$~Hz for a $10^{6} M_{\odot}$ BH).
The rapid non-monotonic variations in the shaded regions shown in
Fig.~\ref{fig:quadrupoleIC} when $j=0$ (left panel) disappear when
$j=0.06$ (right panel), and the integrable orbits get grouped in a band
around $D(\rho)\approx2$, which indicates they are becoming confined 
to a 2-dimensional torus. 
The results presented here are examples of the characteristic
of what we find in numerical experiments over a vast region of the parameter
space that represent different source configurations and initial conditions.

\subsection{Gravitational waves in the semi-relativistic approximation}
While the appearance of the features highlighted in
Fig.~\ref{fig:quadrupoleIC} are considered as generic chaotic signatures,
these quantities are not physical observables. 
As the non-integrability of a Hamiltonian system changes the evolution of the
fundamental frequencies of the orbital 
motion~\cite{Kiuchi:2004bv,Apostolatos:2009vu}, the islands of
instability and prolonged resonant regimes, shown in
Fig.~\ref{fig:quadrupoleIC}, must be encoded directly in the gravitational
waves emitted~\cite{Barausse:2020rsu} by these systems. 

To show how chaos may affect the gravitational waves emitted, we follow
Ref.~\cite{Gair:2007kr} and calculate approximate gravitational waveforms
using the semi-relativistic approximation~\cite{Babak:2006uv},
i.e., neglecting gravitational wave dissipation and the conservative
self-force. 
This semi-relativistic ``kludges" are widely used to explore generic
EMRIs~\cite{Chua:2017ujo, Chua:2020stf}.

\begin{figure*}[htb]
\centering
\includegraphics[width=.6\textwidth,clip=true]{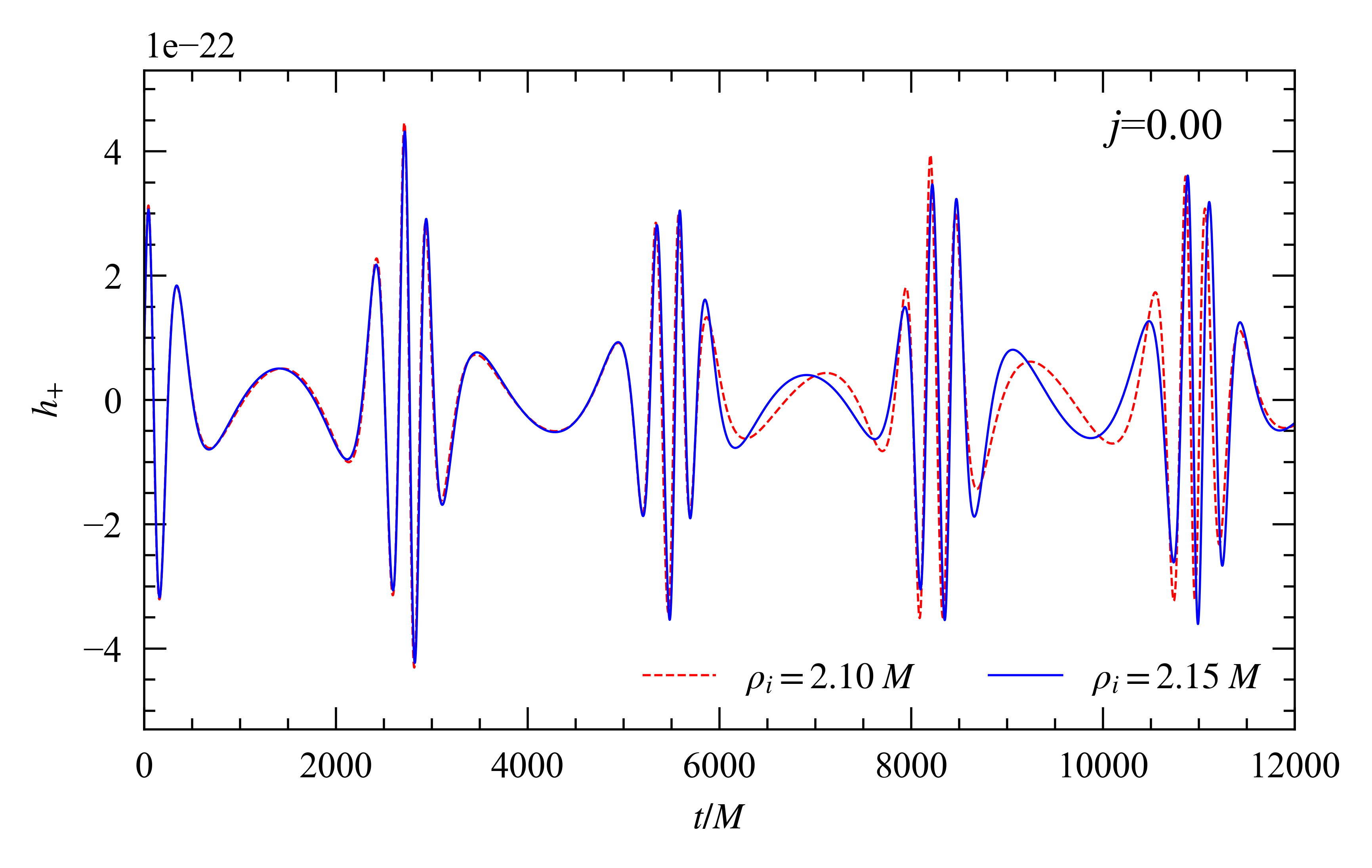}
\caption{Approximate waveforms ($h_{+}$) produced by a 
$20 M_{\odot}$--$10^6 M_{\odot}$ at $4$ Gpc, observed at polar angle $\pi/6$
and azimuthal angle $\pi/4$, for $j=0.0$, for two nearby initial conditions,
i.e., at $\rho_i=2.1\, M$  (red dashed lines) and at $\rho_i=2.15\, M$ 
(blue solid lines). 
Other orbital parameters are the same ones as those used in
Fig.~(\ref{fig:quadrupoleIC}), i.e., $E=0.95$ and $L=3.0\, M$ 
and $\mathcal{Q}^{\rm{G}}=1.6 \, M^3$. As it can be seen, the waveforms
slowly start to differ given the difference in the initial conditions, and
chaos is not easily manifest in the time domain.}
\label{fig:GWS_t}
\end{figure*}

We have evolved two nearby orbits, $\rho_i=2.10\,M$ (denoted in red in
Fig.~\ref{fig:quadrupoleIC}) and $\rho_i=2.15\,M$ (denoted in blue in
Fig.~\ref{fig:quadrupoleIC}) and computed the approximate gravitational
waves. 
In Fig.~\ref{fig:GWS_t} we show the ``plus'' polarization,
$h_{+}$, of these two initial conditions for $j=0.0$. 
While these waveforms are very rich in structure, due to the eccentricity and
inclination of the orbits, both features expected for some EMRIs that will
radiate in the LISA band~\cite{Gair:2017ynp}, the chaotic features are not
manifest in the time domain.

\begin{figure*}[htb]
\includegraphics[width=\columnwidth,clip=true]{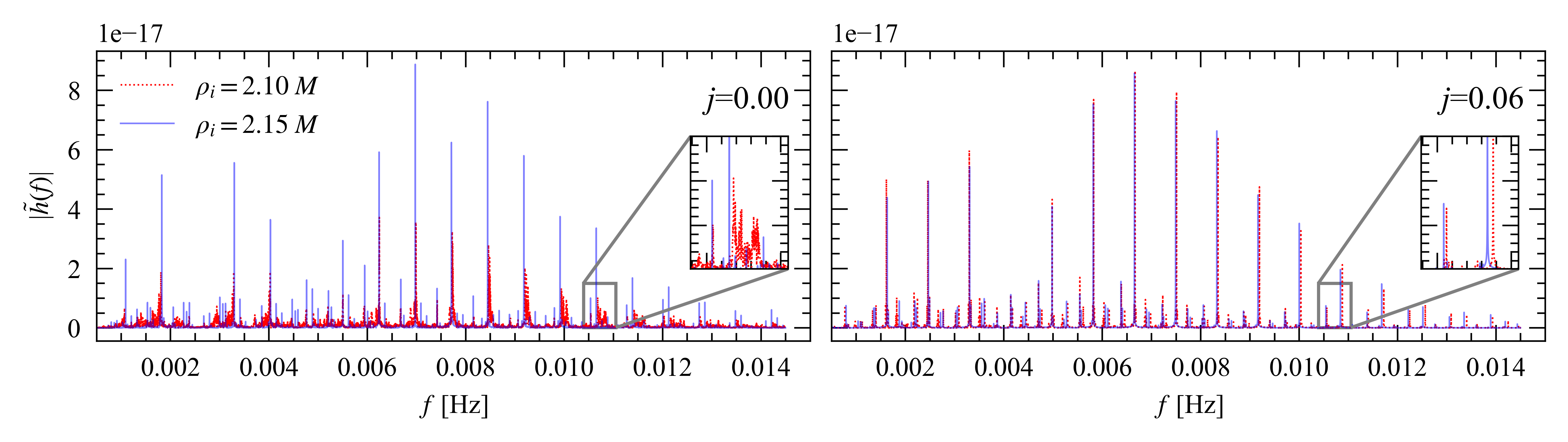}
\caption{Absolute value of the Fourier transform of $h=h_{+}+i h_{\times}$
produced by a $20 M_{\odot}$--$10^6 M_{\odot}$ at $4$ Gpc, observed at
polar angle $\pi/6$ and azimuthal angle $\pi/4$, for $j=0$ (left panel)
and $j=0.06$ (right panel), for two very close initial conditions, i.e.,
at $\rho_i=2.1\, M$  (red dashed lines) and at $\rho_i=2.15\, M$ (blue
solid lines). 
Other orbital parameters are the same ones as those used in
Fig.~(\ref{fig:quadrupoleIC}), i.e., $E=0.95$ and $L=3.0\, M$ 
and $\mathcal{Q}^{\rm{G}}=1.6 \, M^3$. 
Orbital chaos manifests itself through a multi-peak structure in the
Fourier amplitude of gravitational waves, as seen in the
inset of the left panel. 
When the spin-parameter is increased, however, this multi-peak structure
is suppressed, and thus suggesting that the chaotic-to-regular dynamics
transition may be addressed in terms of the frequency content of the
spectrum.
}
\label{fig:GWS}
\end{figure*}

In the frequency domain, however, the separation between regular
and chaotic conditions is more clear.
In Fig.~\ref{fig:GWS} we show the Fourier transform of these approximate
waveforms. 
Observe that when $j=0.0$, (left panel), the chaotic trajectories induce a
multi-peak structure on the Fourier transform of the emitted gravitational
waves, while this structure is absent when $j=0.06$ (right panel). 
The difference between both spectra is now evident: 
gravitational waves from chaotic configurations are denser in frequency
content than gravitational waves from regular configurations, as a result of
the very nature of chaos. 
In the presence of frame-dragging, $j\neq0$, these ``extra" frequency
components induced by chaos are suppressed and the spectra becomes
simpler.

For sufficiently slow evolutions, i.e., small-mass ratios, this
richness and structure of the frequency components may have a measurable
impact in the gravitational waveform~\cite{Apostolatos:2009vu}.  
As we chose the source parameters for the approximate gravitational
waveforms to fall well within the LISA band~\cite{Gair:2017ynp}, for an
astrophysical source this would imply that detecting chaos may be more
difficult than previously anticipated due to the stealth effect that frame-dragging has.

\section{KAM Tori Reconstruction II}
\label{KAMII}
\begin{figure}
\centering
\includegraphics[width=.7\textwidth]{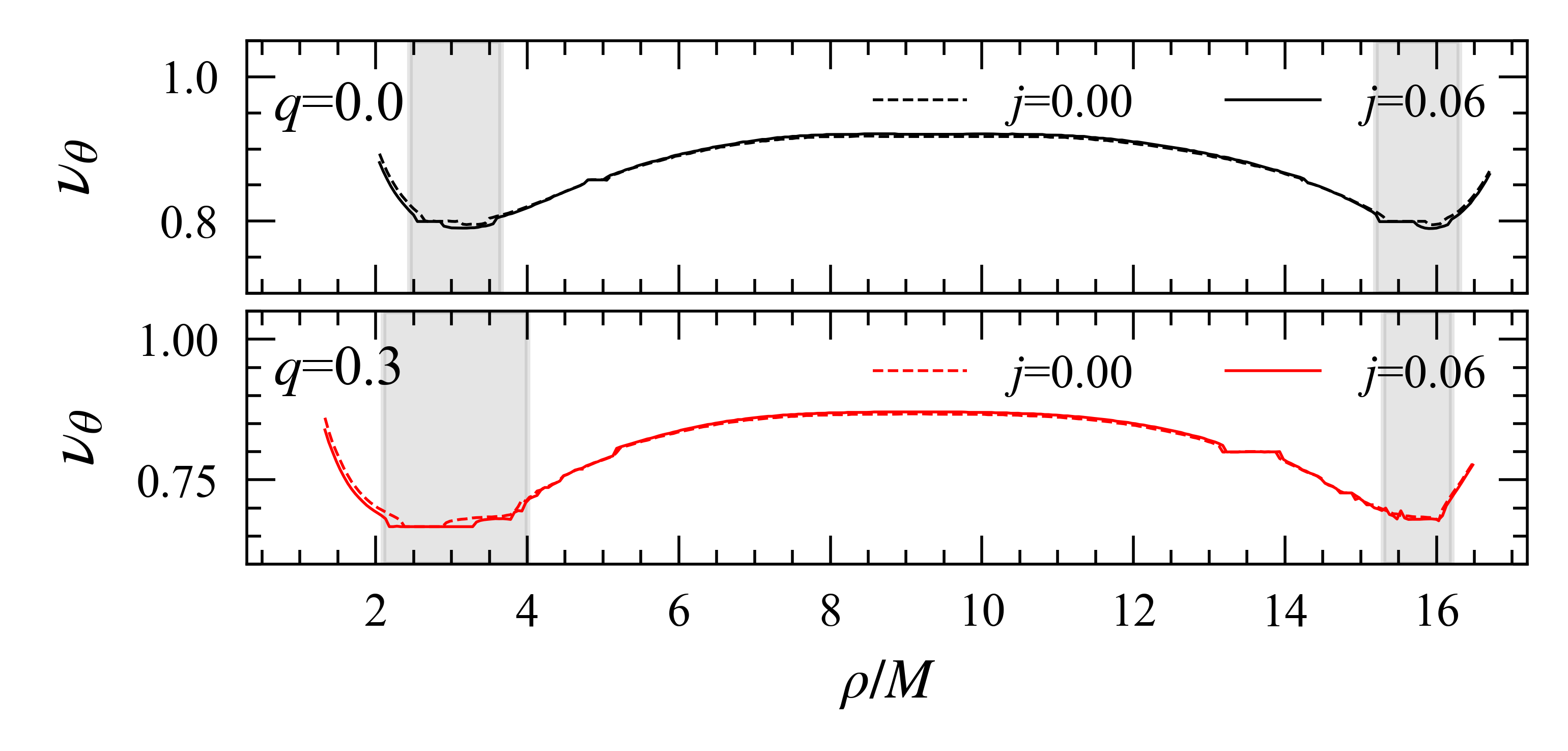}
\caption{Rotation curves for an orbit with $E=0.95$ and $L=2.30\, M$ in a
background with $\mathcal{Q}^{\rm{G}}=0$, $\mu=15 \, M^2$, and $q=0.0$
(top) or $q=0.3$ (bottom), and with dimensionless spin parameter $j=0.0$
(dashed lines) and $j=0.06$ (solid lines), while maintaining
$\mathcal{Q}^{\rm{G}}$ constant. 
As in the previous cases, an increase in the spin parameter generates a
smooth transition to regular dynamics. 
In the charged case, however, spin is not as effective as before at
suppressing chaos.}
\label{Fig:rotsII}
\end{figure}

Let us now study a different example in which the spacetime has an electromagnetic contribution. Let us consider a test particle in an orbit with $E=0.95$ and
$L=2.3\,M$ in a PRS spacetime~\cite{Pachon:2006an}, with mass $M$,
vanishing quadrupole $\mathcal{Q}^{\rm{G}} = 0$ and a magnetic dipole
$\mu=15\, M^2$. 
The electromagnetic multipole moments now break the integrability of the
system, as shown in the shaded regions of the top panel of
Fig.~\ref{Fig:rotsII} by the non-monotonic behavior of the rotation curve.
This is not because of the mass quadrupole moment, which is zero here, but rather due to energy density of
electromagnetic field sourcing a non-trivial mass octupole moment
$\mathcal{H}^{\rm{G}}$ [see Eq.~(\ref{multipolosP3})].
As in the previous case, when the spin parameter is increased from 
$j=0.0$ to $j=0.06$, the chaotic behavior is suppressed, but this time
the suppression is not as effective as in the not magnetized case. 
This is shown in the top panel of Fig.~\ref{Fig:rotsII} by
the length of the plateaus, corresponding to the resonant KAM tori,  
decreasing in the rotating case.

For rotating sources without electromagnetic fields ($\Phi = 0$, as explored in the previous section), $\Im [\mathcal{E}] \neq0$ and provides the standard contribution of the spin angular momentum to the vorticity scalar of the spacetime. 
On the other hand, in the presence of non-aligned electric and magnetic 
fields, the terms $\Phi^{*}\Phi_{,z}$ and $\Phi^{*}\Phi_{,\rho}$ are
non-zero, and therefore the spacetime is endowed with a non-vanishing vorticity scalar. 
The contribution of the electromagnetic fields means that one can generate
spacetime vorticity when $ q \mu \neq 0$, even if the central object is
not spinning, as originally suggested by Bonnor~\cite{BONNOR199123}, and
later explained in Ref.~\cite{Herrera:2006cw}.

The fact that the vorticity scalar can also be sourced through the 
electromagetic field, naturally allows for a new scenario to study the
suppression of chaos.
However, the presence of an electric charge monopole also induces a prolate
contribution to quadrupole deformation, which tends to generate chaotic
dynamics~\cite{Pachon:2010yg}. 
The bottom panel of Fig.~\ref{Fig:rotsII} shows how the electric charge of the source creates new chaotic zones, but as spin is increased from $j=0.0$ to $j=0.06$, these chaotic structure is not suppressed as efficiently as in the previous section.

\section{Is the vorticity of the spacetime suppressing chaos in the PRS metric?}
\label{structure}

To understand the behavior displayed in the examples shown in
the previous two sections, let us consider the complex Ernst potential
$\xi$, which has a direct relation to the Newtonian gravitational 
potential~\cite{Ernst:1967by,Sotiriou:2004ud,Pachon:2010yg} and from which
the multipole moments in Eq.~(\ref{multipolosP3}) were
derived~\cite{Sotiriou:2004ud}. 
By performing a Taylor expansion of $\xi(\rho,z)$ in spherical coordinates
$\left(r,\theta \right)$ at spatial infinity, its real $\Re[\xi]$ and
imaginary $\Im[\xi]$ parts are
\begin{eqnarray}
\Re[\xi] &\approx  -\frac{M}{r} - 
\frac{\mathcal{Q}^{\rm{G}}}{r^3}P_2(\cos \theta) 
+ O(r^{-5})\,, 
\nonumber \\
\Im[\xi] &\approx \frac{aM}{r^2}P_1(\cos \theta)+ O(r^{-4})\,, 
\label{Eq:asympot}
\end{eqnarray}
where $r=\sqrt{\rho^2+z^2}$, $\cos \theta=z/r$ and $P_n$ denotes the
Legendre polynomial of $n$-th degree. 
The real part of the Ernst potential provides information about the
Newtonian gravitational potential~\cite{Sotiriou:2004ud,Pachon:2010yg},
which contains contributions from the mass monopole and the mass quadruple
deformation~\cite{PhysRevE.63.035201,LETELIER20113655}.

Given that Newtonian orbits for two test-particles (with zero
quadrupole moment) are integrable, it is the quadrupole moment that could
break the integrability of the systems~\cite{PhysRevE.63.035201,LETELIER20113655}. 
The imaginary part of the Ernst potential, on the other hand, contains the
leading-order contribution of the rotation of the source to the gravitational
field, and therefore it is responsible for the
reconstruction of the broken tori. 
This is because this term changes the effective potential barrier for the
orbiting particles, which we have kept fixed by not changing $E$
and $L$, so that the contribution of the quadrupole moment
deformation, which is what sources chaos, becomes effectively weaker.
Due to the analytical control of the multipole structure of the
source that we have in this particular spacetime, we are able to uniquely vary the magnitude of the vorticity scalar (and thus, the amount of frame-dragging) while keeping the mass multipoles $M$ and $\mathcal{Q}^{\rm{G}}$ constant,
regardless of the value of the angular momentum of the source.
Thus, the larger the spin angular momentum, the stronger the contribution of
this term to the dynamics and therefore, the smaller the deviations from the
Kerr spacetime. 

Given that deviations from the Kerr solutions are not unique, increasing the vorticity scalar of the spacetime may not always display this suppression, because of other more dominant effects. For example, the fact that the suppression of chaos in the electromagnectic case is less effective happens because the electric charge monopole plays two roles simultaneously: it induces a spacetime vorticity scalar (and therefore
frame-dragging), which suppresses chaos, but at the same time, it also
induces a more prolate deformation, which sources chaos. 
Our numerical explorations indicate that the suppression effect is smaller than the sourcing of chaos, and that overall, vorticity-induced chaos suppression is less effective when the source is charged than in the uncharged case.  

\section{Concluding remarks}
\label{Future}
We have investigated extreme mass-ratio inspirals in the PRS bumpy
spacetime through a test-particle approximation and found a novel
mechanism, i.e., frame-dragging-assisted suppression of chaos. 
The analytical control of the multipole structure of the PRS
metric, allowed us to introduce frame-dragging effects while keeping the
quadrupole moment $\mathcal{Q}^{\rm{G}}$ of the source fixed, and see its
effect to the dynamics of the motion of the test particles. 
Given that the quadrupole moment is responsible for the break of the
integrability of the system, frame dragging (as quantified through the vorticity scalar) increases the regularity of the
orbits by making its contribution weaker.

A natural extension of the work presented here is towards
studying the motion of charged particles, albeit not in the context of
EMRIs. 
When considering \emph{charged} particles, previous works have not found any
clear and unique indication of the spin dependence with chaos (see for
instance, Refs.~\cite{Takahashi:2008zh, Kop_ek_2010}).
Note that these results are \emph{not} in conflict with our findings, as
we are considering here \emph{neutral} test particles and showed explicitly
how the spin contribution of the source affects the multipole structure of
the source and interacts with the particle. 
This effect may also be presented in such cases, but perhaps subdominant with
respect to electromagnetic forces, and therefore not easily detectable.

As gravitational waves probe both the conservative
(time-symmetric) and  the dissipative (time-asymmetric) sectors of the
gravitational theory, neglecting radiation reaction is only valid as a first
approximation towards the understanding of the characterization of chaotic
features with EMRIs. 
Despite recent impressive calculations from black-hole perturbation theory
(see, for instance, Ref.~\cite{Pound:2019lzj}), the EMRI modeling within GR
is still not finished. 
This makes the addition of a consistent radiation reaction for this setup out
of the scope of this work.

The suppression discussed in this work implies that, although chaos may be
technically present in the orbital motion and the gravitational waves
emitted by extreme mass-ratio inspirals around non-Kerr central objects, 
it may be hidden by frame-dragging effects. 
If chaos operates in a stealth mode for orbits around spinning objects, its
detectability (or constraints on its existence) may be challenging with
future gravitational wave observations. 

\section*{Acknowledgments}
We thank the referees for their feedback, which enabled us to
improve the presentation of our results and manuscript.
A.F.G.-R. and L.A.P acknowledge financial support through Banco de la
Rep\'ublica grant 201926 (Project 4345).
A.C.-A. and N.Y. acknowledge financial support through NASA grant
No.~NNX16AB98G, No.~80NSSC17M0041, No.~80NSSC18K1352 and NSF grant
PHY-1759615. 
A.C.-A. also acknowledges funding from the Fundaci\'on Universitaria Konrad
Lorenz (Project 5INV1). 
Computational efforts were performed on the Illinois Campus Cluster, a
computing resource that is operated by the Illinois Campus Cluster Program
(ICCP) in conjunction with the National Center for Supercomputing
Applications (NCSA) and which is supported by funds from the University of
Illinois at Urbana-Champaign.
\section*{References}
\bibliographystyle{iopart-num}
\bibliography{references.bib}

\end{document}